\documentstyle[aps,prl,multicol,epsfig]{revtex}
\begin{document}

\title{Dynamical Mean Field Theory of the Antiferromagnetic Metal to
Antiferromagnetic Insulator Transition}
\author{ R. Chitra and G. Kotliar}
\address
{Serin Physics Laboratory, Rutgers University,
Piscataway, NJ 08855, USA}
\date{\today}
\maketitle
\begin{abstract}
We study the antiferromagnetic  metal  to antiferromagnetic insulator
using dynamical mean field theory and exact diagonalization
methods.  We find two qualitatively
different behaviors depending on the degree of  magnetic
correlations.  For strong correlations combined with
magnetic frustration, the transition can be
described in terms of a renormalized slater theory, with
a continuous gap closure driven by the magnetism  but strongly
renormalized by correlations. For weak magnetic correlations, the transition
is weakly first order.


\end{abstract}

\begin{multicols}{1}

The  correlation driven metal insulator transition (MIT) or Mott transition 
is one of the central problems of condensed matter physics.
For generic band structures, it occurs at a finite value of the interaction
strength and constitutes  a genuinely non perturbative phenomena.
Recently, a great deal of progress has been made using
the Dynamical Mean Field approach (DMFT),
 a method which becomes exact
in the well defined limit of infinite lattice coordination \cite{metzner,rmp}.
However, all  studies  so far, have been  confined to the
paramagnetic  metal (PM)  to paramagnetic insulator (PI) transition.
In this letter, we study the  transition from the 
antiferromagnetic metal  (AM) 
to an antiferromagnetic insulator  (AI) within DMFT.
The motivation for this work is twofold. Experimentally,
the  interaction or pressure driven
MIT in $V_2 O_3$\cite{V2O3,rev} and $Ni S_{2-x} 
Se_x$ \cite{NiS,rev}
take place between {\it magnetically ordered } states. While
we are still quite far from a realistic modeling of these oxides,
(in this work we only consider commensurate magnetic order
and ignore orbital degeneracy and realistic band structure)
we would like to take into account the effects of magnetic long
range order present
in these systems. On
the theoretical side, we would like to understand how
magnetic correlations , which controls the scale at which the spin
entropy is quenched, affect the MIT.

We consider the Hamiltonian,

\begin{equation}
H=t_1 \sum_{nn} c^{\dagger}_i c_j +t_2 \sum_{nnn} c^{\dagger}_i
c_j + U\sum_i n_{i\uparrow}n_{i\downarrow} - J\sum_{nn} {\bf S}_i.
{\bf S}_j
\label{ham}
\end{equation}
\noindent
$t_1$ and $t_2$ are nearest and next nearest neighbor hoppings respectively and
$U$ is the onsite coulomb repulsion. 
J is a  ferromagnetic spin spin interaction that
we add to the model in order to  independently control  the
strength of the antiferromagnetic correlations of the system
\cite{laloux}.
We parametrise the hoppings as  $t_1^2 = (1-\alpha) t^2$ 
and  $t_2^2 = \alpha t^2$ where $\alpha$ is the degree of frustration. 
 $t_2$  breaks the perfect nesting characteristic of the bipartite
lattice 
\cite{Marcelo,Moreo} and allows for a direct AM to AI transition.

Within the dynamical mean field approximation, in the presence of 
magnetic order, the single particle Green's function
on the $A$ and $B$ Neel sublattices has the following form
\begin{equation}
G_\sigma^{-1}=\pmatrix {
i\omega_n +\mu_\sigma -{\tilde \epsilon}_k -\Sigma_{A\sigma} & -\epsilon_k
\cr
-\epsilon_k  & i\omega_n +\mu_{-\sigma}-{\tilde \epsilon}_k -\Sigma_{B\sigma} 
 \cr}
\end{equation}
\noindent
$\sigma=1,2$ refer to the spin up and down states, $m$  the
staggered
magnetisation 
and
$\mu_\sigma=(\mu -{U\over 2}+(-1)^\sigma Jm)$. 
 $\Sigma$ are the self energies which are momentum
independent within
this approximation.  
The  local  Green's functions  $G_{ii\sigma}$ depend on the sublattices
(A and B) and obey 
$G_{A1}=G_{B2}\equiv G_1$, 
$G_{A2}=G_{B1}\equiv G_2$. Similarly,  
$\Sigma_{A1}=\Sigma_{B2}\equiv \Sigma_1$ and 
$\Sigma_{A2}=\Sigma_{B1}\equiv \Sigma_2$. 

Within DMFT, these local self-energies  and  Greens functions
are obtained  from  an  Anderson impurity model
\begin{eqnarray}
H_{imp} =&&\sum_{l\sigma} \epsilon_{l\sigma} d^{\dagger}_{l\sigma}
d_{l\sigma}+ \sum_{l\sigma} V_{l\sigma}(d^{\dagger}_{l\sigma}f_{\sigma} 
+h.c.) \\ \nonumber
&& -\mu_\sigma  \sum_\sigma f^{\dagger}_{\sigma} f_{\sigma}
+U n_{f\uparrow}n_{f\downarrow}
\label{himp}
\end{eqnarray}
\noindent
 describing  an impurity electron $f$ hybridising
with a bath of conduction electrons $d$
via the spin-dependent hybridisation functions 
\begin{equation}
\Delta_\sigma(i\omega_n)= \sum_l {{V_{l\sigma}^2} \over {i\omega_n -\epsilon_{l
\sigma}}} 
\label{hybd}
\end{equation}
\noindent
where $\omega_n$ are the Matsubara frequencies.
The $\Delta_\sigma$ obey  self consistency
conditions which can be expressed in terms of $G_1$, $G_2$ and the
non-interacting density of states \cite{rmp}.
We restrict
 ourselves   to the Bethe lattice in the following.
The non-interacting density of states on the Bethe lattice  
is given by 
$D(\epsilon)= {2\over {\pi D^2}}\sqrt{D^2 -\epsilon^2}$ with $D={\sqrt 2}t$ 
being the half bandwidth.
The  self-consistency conditions now have a simple form
\begin{equation}
\Delta_\sigma (i\omega_n)= {{t_1^2} \over 2} G_{-\sigma}(i\omega_n)
 + {{t^2_2} \over 2} G_{\sigma}(i\omega_n)
\label{sc}
\end{equation}
\noindent
 We consider only the half-filled case, which on the Bethe
lattice is achieved when 
$\mu$  is chosen to be $U/2$ due to a special
particle hole symmetry of  the
 impurity problem :
$d^\dagger_{l\sigma} \to d_{l-\sigma}$,  
$d_{l\sigma} \to d^\dagger_{l-\sigma}$ , 
$f_{\sigma} \to f^\dagger_{-\sigma}$,  
$f^\dagger_{l\sigma} \to f_{-\sigma}$ and with $(\epsilon,V)_{l\sigma} \to
-(\epsilon,V)_{l-\sigma}$ provided $\mu = U/2$.  
    
To study  these equations   
 we use   the algorithm of Ref.\onlinecite{Krauth}
which uses zero temperature Lanczos method
 to compute the Green's functions of the impurity model given by (\ref{himp})
and iterate the model until the self-consistency given by 
(\ref{selfcon}) is achieved.
Though the results to be presented  were obtained for  the case where the bath
was represented by $5$ sites,  several of them  were checked  for 
$N=7$ to establish the results seen for $N=5$. 
 The typical number of iterations
varied between  $50-65$ for robust convergence.
We found substantially different behavior in the 
cases of  strong and weak correlations (induced by a quenching of antiferromagnetism
 mimicked by ferromagnetic interactions in this paper) which we discuss separately.

\subsection {Strong Magnetic Correlations}
For small U, the system is in the paramagnetic phase.  
As $U$ increases beyond  a critical $U_{cm}$, antiferromagnetic moments
develop and the up and down spectral functions  $\rho_{\sigma}(\omega) =
(-1/\pi){\rm{Im}} G_\sigma (\omega)$
are no longer
equal. The spin up spectral
function has more spectral weight in the upper Hubbard band than in the
lower Hubbard band  and vice versa for the down spin spectral function.
The low frequency peak in $\rho_\sigma(\omega)$ is no longer
centered 
around $\omega=0$ but is split into two peaks  centered around
some $\pm \omega_0$ with a minimum at $\omega=0$. This can
be attributed to the fact that the effective staggered magnetic
field that is generated when antiferromagnetism develops splits
the quasiparticle bands.
As $U$ is increased further, the density of states at $\omega=0$
decreases until a critical $U_{MIT}$ where $\rho_{\sigma}(\omega=0)
=0$ and the system becomes insulating. For $U \ge U_{MIT}$,
a gap $\Delta$  which grows with $U$, opens continuously. 

\begin{figure}
\centerline{\epsfig{file=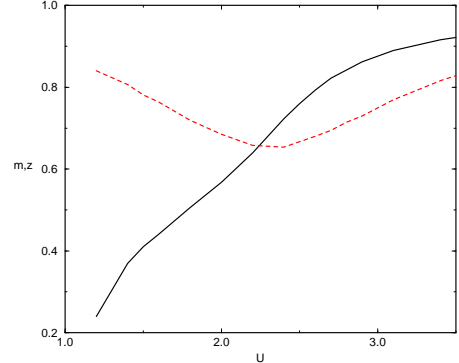,angle=270,width=7cm}}
\narrowtext
\caption{The magnetization $m$ (bold line)  and $z$ (dashed line)
 versus U for $t_2/t_1= 0.57$ with $U_{MIT}=3.0t$. 
}
\label{fig:mzj0}
\end{figure}

A plot of the staggered magnetization $m$ versus $U$ is shown in
Fig. \ref{fig:mzj0}. $U$ is in units of $t$ in all the graphs.
$m$  increases monotonically with $U$ 
and does not exhibit any special feature at the MIT. 
Note that though the magnetic moment is quite large at
the MIT  and increases with increasing $t_2/t_1$   
it   saturates  only when
one is well into the insulating phase. 
In the presence of magnetic long range order,
the  self energy does not
show any anomalous behavior across the MIT. This should be contrasted
with the 
paramagnetic case where ${\rm Im}\Sigma$ diverges as $(i\omega_n)^{-1}$
at the MIT.
To obtain an  insight into the nature of the MIT,  
we use the following low energy parametrization of
 $\Sigma_1 $ and
$\Sigma_2$ in conjunction with the numerical results
\begin{eqnarray}
\label{sig}     
\Sigma_1(i\omega_n) &= &h + (1- {1 \over {z}}) i\omega_n \\ \nonumber
\Sigma_2(i\omega_n) & =& -h + (1- {1 \over {z}}) i\omega_n
\end{eqnarray}
\noindent
where  $h$ is the staggered field generated by the interactions in the
antiferromagnetic phase.
$z$ goes to zero at the PM-PI transition, but
in the presence of magnetism  it  is non-monotonic and remains non-zero 
even in the insulating phase (cf. Fig.\ref{fig:mzj0}). 
\noindent
In terms of the local Green's functions, $\Sigma_\sigma$ is
determined by the following expression on the Bethe lattice
\begin{equation}
{G}_{\sigma}^{-1}+ \Sigma_\sigma= i\omega_n + \mu_\sigma
 -{{
 t_1^2} \over 2} G_{-\sigma}
-{{t_2^2} \over 2} G_{\sigma}
\label{selfcon}
\end{equation}
\noindent
Rewriting  (\ref{selfcon}),
we obtain
\begin{eqnarray}
\label{crit}
1&=& (i\omega_n + {\rm Im}\Sigma_1
 +{\rm Re}\Sigma_1 ) G_1 -{{t_1^2} \over 2} G_1 G_2 - {{t_2^2} \over 2} G_1^2 \\ \nonumber
1&=& (i\omega_n+  {\rm Im}\Sigma_2 +{\rm Re} \Sigma_2
 ) G_2 -{{t_1^2} \over 2} G_1 G_2 - {{t_2^2} \over 2} G_2^2 
\end{eqnarray}
\noindent
Using (\ref{sig}) in (\ref{crit}) and continuing to real frequencies
 we obtain the low frequency spectral
functions
\begin{equation}
\rho_{1,2} (\omega)= 
\rho_(0)[ 1\pm {{\omega} \over {
hz}} ] 
\label{spec}
\end{equation}
\noindent
where $\rho(0)$ is the density of states at the Fermi level and is given by
\begin{equation}
\rho(\omega=0)= {2 \over { \pi D^2}} \sqrt{D^2 - ({h \over { \alpha}})^2}
\label{rho}
\end{equation}
\noindent
Note that $\rho(0)$ does not depend on $z$.
Since $h$ increases monotonically with $U$, 
$\rho(0)$ decreases with increasing $U$ in the magnetically ordered phase.
This yields a critical value $h_c$ where $\rho(0)\to 0$
 and a gap opens continuously.  
 This is the point at which the MIT occurs.
We, therefore, see that the MIT in this case is of a very different
nature and there are no Kondo resonances which disappear 
discontinuously at the MIT.
Our results indicate that   
 in the vicinity of $U_{MIT}$,  $h$   increases
linearly  with $U$ implying that $\rho(0)$ vanishes as $ (U_{MIT} -U)^
{\frac1{2}}$ as we approach the transition from the metallic side.

\begin{figure}
\centerline{\epsfig{file=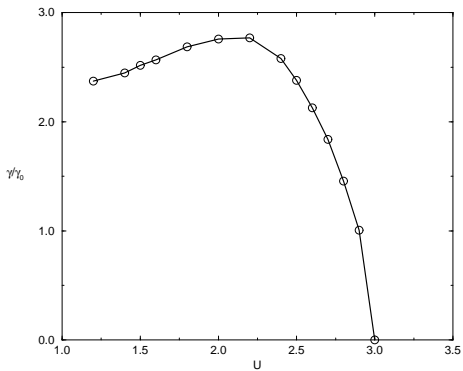,angle=270,width=7cm}}
\caption{ $\gamma/ \gamma_0$ vs.  $U$
for  
 $t_2/t_1=0.57$ ($U_{MIT}=3.0t$). $\gamma_0$ has been defined in the text.} 
\label{fig:ngam}
\end{figure}

We now turn to  the behavior of the
coefficient of the linear term of the specific heat   $\gamma$.
The basic physics controlling the behavior of the  specific heat
in the antiferromagnetic phase
is the
competition between  the
 the increase of the effective mass
$  m^* $
and the decrease of
the  density of states $\rho(0)$.
Though $ m^*  \propto  [1- {{\partial \Sigma(i \omega_n)}
\over {\partial i \omega_n}}] = z^{-1}$,
initially increases, its increase
is  cut off by the staggered magnetization (cf. Fig. \ref{fig:mzj0}).
When the  staggered magnetization $m$ is
large  the decrease in $\rho(0)$ is the dominant effect,
whereas the first effect
dominates when the magnetism is weak.
The latter is true in the case of the paramagnetic MIT  
 where $m^*$ and hence $\gamma$ are both $\propto z^{-1}$  
and diverge as $z \to 0$ at the transition.
  This divergence is
 related to the fact that there is a residual entropy in the insulating
 phase which results from the spin degeneracy at every site.
 However, such a degeneracy is lost in the case where the insulator is
 also an antiferromagnet. Infact using Eqs. (\ref{sig},\ref{spec},\ref{rho}) we find
\begin{equation}
{\gamma \over {\gamma_0}} =\frac 1{z} {\sqrt{ D^2 -
 {{h^2} \over {\alpha^2}}}}
\label{dos}
\end{equation}
\noindent
where $\gamma_0 =  {{\pi k_B^2} \over {3t}}$ is the specific heat
coefficient of the non-interacting problem.  
Since $z$ remains nonzero,  
  $\gamma\to 0$ as $(U_{MIT} -U)^\frac1{2} $ at the MIT. 
As anticipated, we see in  Fig.\ref{fig:ngam} that $\gamma$ increases with $U$ for small
$m$ 
and decreases
 for larger moments.  

We can generalize the above to a lattice with a realistic dispersion 
$E_{\bf k}= \epsilon ({\bf k}) +{\tilde \epsilon}({\bf k})$
where  $ \epsilon$ and $\tilde \epsilon$ are the contributions from the
$t_1$ and $t_2$  hoppings.
The Fermi surface of the non-interacting system is defined by
$ E_{{\bf k}_F} =0$.                                  
When the interaction $U$ is turned on, the dispersions of the
\noindent
quasiparticles change and are given  by the poles of
the Green's function given by Eq. 2.
From the preceding analysis, we can  say that
the main effect of interactions is to modify the band dispersion
in the following manner
\begin{equation}
E_{\bf k}= z[
{\tilde \epsilon}_{\bf k} \pm \sqrt{ \epsilon_{\bf k}^2
+ h^2}]
\label{band}
\end{equation}
\noindent
In the non-interacting case ($h=0$ and $z=1$)   the two bands  overlap and
there is no gap at the Fermi surface.
For small $U$ in the paramagnetic
phase there is no gap in spectrum but the bands are slightly
renormalized by the factor $z$. When antiferromagnetic
moments set in, the non-zero $h$ changes the band curvature
and the renormalized bands start
moving away from each other reducing the Fermi surface area.
 The manner in which the Fermi surface area decreases is
described 
 by the density of states $\rho(0)$.
 For small $h$   
 there are 
regions where the bands overlap and there is no gap in the system.
At a critical value  $h=h_c$ 
 a gap opens up in the spectrum as  the Fermi surface shrinks
to zero signalling the MIT.   

\subsection{Weak Magnetic Correlations: $J\ne 0$}
The previous  scenario   was characterized by relatively weak
electron correlations, the quasiparticle residue $ z$ near the transition,
was at most  $0.65$, indicating that a relatively
high fraction of the spectral weights remain coherent.
This can be understood in simple qualitative terms,
when the magnetic frustration is weak, the relatively
large magnetic exchange produces  a large magnetization,
which in turn reduces the double occupancy and hence the
correlations. All the spin entropy is quenched by the spin ordering,
and the presence of a hopping in the same sublattice favors 
coherent quasiparticle propagation.
To access the strongly correlated regime  where most of
the spectral function is incoherent we 
we turn on a nonzero $J$ 
which reduces $h$ and $m$ and
hence increases $U_{MIT}$ as suggested by
(\ref{rho}). 

\begin{figure}
\centerline{\epsfig{file=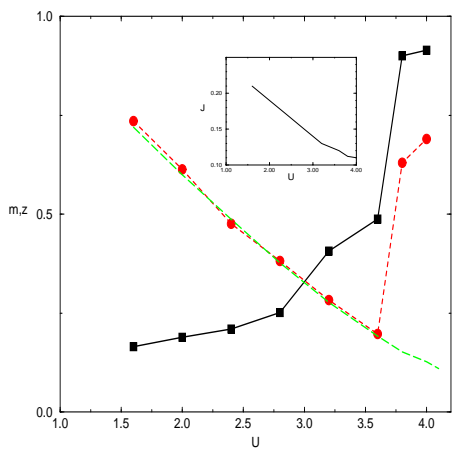,angle=180,width=7cm}}
\caption{ $m$ (bold line) and $z$ (dashed line) vs.  $U$
 for the case of small frustration $t_2/t_1=0.2$ with $J$  varying as
a function of $U$. $U_{MIT}= 3.8t$ in this case.
 The inset shows the $U$ dependence of $J$.  
} 
\label{fig:mzjofu}
\end{figure}

We studied the case $t_1^2= 0.95 t^2$ and $t_2^2= 0.05 t^2$ with $t_2/t_1=
0.22$.  For $J=0$, this
system  shows the usual band MIT discussed in the previous section,
 in the vicinity of $U=2t$. 
By choosing a $J(U)$ cf. inset of Fig.\ref{fig:mzjofu}, such that the magnetic moment remains very small
for a large regime of $U$ as shown in Fig. \ref{fig:mzjofu}, we 
move the MIT 
 to $U=3.8t$. 
In the antiferromagnetic
metallic phase, a Kondo like resonance persists in the spectral function
 up to $U=3.8t$
and disappears suddenly at the transition. This resonance at small $\omega$
splits into two peaks when the magnetic moment becomes appreciable 
as was seen earlier. However, unlike in the paramagnetic case the
height of this resonance is not pinned at the value of the 
non-interaction density of states but decreases with increasing $U$.
We also find that the slope $z$ decreases to values
much smaller than  was seen in the band transition picture 
as seen in  Fig. \ref{fig:mzjofu}. $z$  decreases
until the moment becomes sufficiently large as to stop its reduction. In the
insulator, which still resembles the band-insulator in that there is
a pseudogap, the self-energy is still analytic and one can still define
a $z$.
\noindent
 We find that there is a jump in $z$ and $m$ at the transition in addition to
the discontinuous disappearance of the resonance mentioned earlier,
reminiscent of a
 first order transition.
On re-tracing the insulating solution as function of decreasing $U$, we 
find that there is co-existence. The insulator survives up to $U=3.4t$ and
there is a MIT of the pure band kind at that point and the solution goes
over to one which is a metallic antiferromagnet but with a much higher
value of the moment $m$ and a much larger $z$. This metallic solution
disappears around $U=2.4t$. 
\begin{figure}
\centerline{\epsfig{file=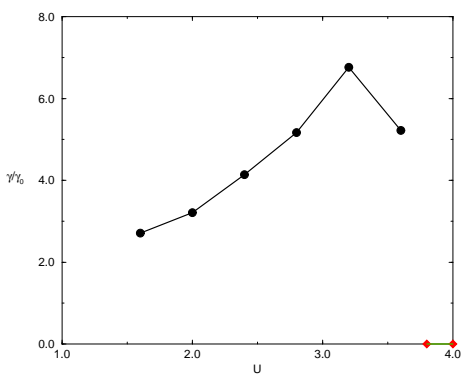,angle=270,width=7cm}}
\caption{ $\gamma$ vs. $U$
for the case of small frustration $t_2/t_1=0.2$ with $J$  varying as
a function of $U$ ($U_{MIT}= 3.8t$).
} 
\label{fig:gam}
\end{figure}
We have measured the double occupancies in
both these solutions and find that the metal with the smaller moment 
has higher double occupancy. This in conjunction with the
 ferromagnetic coupling $J$ ensures that the metal with the lower moment
and higher double occupancy 
is the lowest  energy solution.
In this weak magnetic correlation regime, we find that the specific heat coefficient
$\gamma$ is strongly enhanced cf. Fig.\ref{fig:gam}. This scenario is reminiscent of 
the behavior seen in $V_2O_3$ \cite{V2O3}.     
Note that at the MIT, there is a finite gap which is much smaller than the
Hubbard gap. Also there survive coherent quasiparticle peaks albeit with
very small weight. These coherent peaks are asymmetric about
$\omega=0$ in both $G_1$ and $G_2$.
Close to $U= 4.1t$ these peaks vanish completely and one is left only
with the incoherent Hubbard band structures.
In the coexistent solution, we see that as $U$ is reduced the gap closes
continuously and the the weight of these two coherent peaks increases
continuously too.

To summarize, the character of the  AM-AI
transition is very different from the PM-PI. We found two clearly
distinct scenarios depending  
on the strength of magnetic correlations and frustration in the system.
In the limit of strong magnetic correlations, the transition takes place
as a renormalized Slater transition.  That is, up to a multiplicative
factor which remains finite at the transition,  a gap opens
continuously,
and $\gamma \to 0$ {\it vanish}
continuously , as the MIT is approached
from the metallic side.
On the other hand when development of antiferromagnetic moments in the
metallic phase is suppressed 
a new scenario emerges:
the gap from  the insulating side remains finite at the MIT,
and a substantial increase of the specific heat is observed when the
 MIT is approached from the metallic side.
Our results indicate
that   the transition is
weakly first order.
These two scenarios, are remarkably similar to the observations
reported in $NiS_{2-x}Se_x$ \cite{NiS,NiS2} and $V_{2-y}O_3$ \cite{V2O3}
 respectively and could be developed further to interpret other
experimental observations \cite{rev}.

\end{multicols}

\end{document}